\documentclass{IEEEojcsys}

\usepackage[colorlinks,urlcolor=blue,linkcolor=blue,citecolor=blue]{hyperref}

\usepackage{color,array}

\usepackage{graphicx}

\jvol{XX}
\jnum{XX}
\paper{XXXXXXX}
\pubyear{2025}
\receiveddate{XX XX XX}
\accepteddate{XX XX XX}
\publisheddate{XX XX XX}
\currentdate{31 March 2025}
\doiinfo{OJCSYS.2025.Doi Number}

\newcommand{\T}{\mathrm{T}}
\newcommand{\epsc}{\epsilon_\mathrm{c}}
\newcommand{\epsd}{\epsilon_\mathrm{d}}

\usepackage{cite}
\usepackage{amsmath,amssymb}
\usepackage{mathtools}
\usepackage{amsthm}
\newtheorem*{remark}{Remark}
\usepackage{caption}
\usepackage{subcaption}
\usepackage{algorithm}
\usepackage{algpseudocode}
\usepackage{multirow}
\usepackage{graphicx}
\usepackage{booktabs}
\usepackage{tikz}
\usepackage{pgfplots}
\pgfplotsset{compat=1.18}

\begin{document}

\sptitle{Article Category}

\title{Integrating Reinforcement Learning and Model Predictive Control with Applications to Microgrids} 

\editor{This paper was recommended by Associate Editor XXX. XXXX.}

\author{Caio Fabio Oliveira da Silva\affilmark{1} }

\author{Azita Dabiri\affilmark{1}}

\author{Bart De Schutter\affilmark{1}}

\affil{Delft University of Technology, Delft, Mekelweg 2, 2628CD, The Netherlands}

\corresp{CORRESPONDING AUTHOR: Caio Fabio Oliveira da Silva (e-mail: \href{mailto:c.f.oliveiradasilva@tudelft.nl}{c.f.oliveiradasilva@tudelft.nl})}
\authornote{This research has received funding from the European Research Council (ERC) under the European Union’s Horizon 2020 research and innovation programme (Grant agreement No. 101018826 - CLariNet).}

\markboth{PREPARATION OF PAPERS FOR IEEE OPEN JOURNAL OF CONTROL SYSTEMS}{Caio Fabio Oliveira da Silva {\itshape ET AL}.}

\begin{abstract}
This work proposes an approach that integrates reinforcement learning and model predictive control (MPC) to solve finite-horizon optimal control problems in mixed-logical dynamical systems efficiently. Optimization-based control of such systems with discrete and continuous decision variables entails the online solution of mixed-integer linear programs, which suffer from the curse of dimensionality. 
Our approach aims to mitigate this issue by decoupling the decision on the discrete variables from the decision on the continuous variables.
In the proposed approach, reinforcement learning determines the discrete decision variables and simplifies the online optimization problem of the MPC controller from a mixed-integer linear program to a linear program, significantly reducing the computational time.
A fundamental contribution of this work is the definition of the decoupled Q-function, which plays a crucial role in making the learning problem tractable in a combinatorial action space. We motivate the use of recurrent neural networks to approximate the decoupled Q-function and show how they can be employed in a reinforcement learning setting. 
Simulation experiments on a microgrid system using real-world data demonstrate that the proposed method substantially reduces the online computation time of MPC while maintaining high feasibility and low suboptimality.
\end{abstract}


\maketitle

\section{INTRODUCTION}

\subsection{Motivation}

Complex infrastructure systems, such as energy, transportation, and water networks, are pervasive in our modern world. Analysis and control design of such systems is very challenging due to their size and intricate behavior.
Modeling of such systems often require the combination of discrete and continuous decision variables to adequately capture the system dynamics and operational constraints.
In this context, several modeling approaches \cite{parisio2014model, liu2022modeling, ocampo2008suboptimal} have considered hybrid systems to represent these critical infrastructure networks. 
For instance, the solution of the optimal dispatch problem in energy networks consists of scheduling both the power flow directions and intensities among each of the agents of the network. Some operational constraints require the addition of binary variables for modeling, see \cite{parisio2014model}.
As a result, discrete and continuous decision variables have to be planned over a future horizon to reduce the operation cost considering market conditions. 

For the control of hybrid systems, model predictive control (MPC) arises as a promising technique due to its ability to handle constrained complex systems with discrete and continuous decision variables \cite{borrelli2017}. 
Furthermore, MPC has solid theoretical foundations with regard to stability, performance, and safety \cite{rawlings2017mpcbook}. However, these advantages often come at the price of intense online computational requirements, limiting the use of MPC due to hardware constraints or execution time limitations. This bottleneck is especially more pronounced in MPC for hybrid systems, where the optimizer has also to consider a sequence of discrete decision variables over the given prediction horizon. In this case, the optimization problem then becomes mixed-integer program, which is NP-hard \cite{borrelli05dpforhs}. 

The most widely used technique to solve mixed-integer program in modern solvers is branch-and-bound \cite{richards2005mixed}. In the worst case, the solver has to find solutions for the relaxed problems for each possible combination of the discrete decision variables. This is not scalable with respect to the number of discrete variables because of the combinatorial nature of the problem. Branch-and-bound algorithms mitigate this issue by efficiently pruning branches of the search tree by estimating lower and upper bounds of the objective function and by proposing cutting planes, which reduce the feasible set of the mixed-integer programs by introducing linear inequalities as additional constraints \cite{conforti2014integer}. Moreover, many expert-designed heuristics are employed for improving the search, such as node selection and branch variable selection techniques to reduce computation time. One alternative to branch-and-bound for control of hybrid systems is to pre-compute the MPC control law offline and simply evaluate this function online. For hybrid systems, the offline computation of the explicit MPC control law via multi-parametric programming and dynamic programming was explored in \cite{borrelli05dpforhs}. However, this approach, also referred to as explicit MPC, can be only successfully applied to low-dimensional linear systems. Despite improvements in branch-and-bound and explicit MPC, solving mixed-integer programs with a significant number of integer decision variables remains fundamentally difficult.

Recently, supervised learning has been explored in several MPC approaches for hybrid systems to reduce the online computational time of the resulting mixed-integer programs \cite{masti19_warmStart, masti2020, cauligi21_CoCo, cauligi22_PRISM}. In essence, the aforementioned supervised learning methods have the same structure and learning setting and the main difference lies in the choice of the classifier. These works employ supervised learning to approximate the mapping from the system state to the discrete optimal solution. By applying this classifier to predict the discrete optimization variables, the mixed-integer program is then simplified to an optimization problem consisting only of real-valued variables, significantly reducing the online computational burden of the MPC controller. To build the training dataset, these methods rely on branch-and-bound to solve the control problem -- a mixed-integer program -- to optimality several times. As a result, the main computational issue is sidelined to the offline phase of the algorithm, where typically more computing resources are available. Moreover, in supervised learning, the goal is to reduce the classification error, e.g., the distance between the predicted discrete sequence and the optimal discrete sequence. 
Even though the classification error is typically a suitable proxy for control performance, this might not necessarily be true depending on the mapping from the input sequence to the control objective function.
For a more complete overview of the literature on the intersection of learning and control for hybrid systems, the reader is referred to Section \ref{sec:related_work}.

\subsection{Contributions}

We propose an integrated reinforcement learning (RL) and MPC method that solves mixed-integer linear programs with low computational footprint, low optimality gap, and high feasibility rate.
We build on the existing idea of decoupling the decision on the discrete and continuous variables with learning and MPC \cite{masti19_warmStart, masti2020, cauligi21_CoCo, cauligi22_PRISM}. 
However, we explore a novel paradigm by employing reinforcement learning in place of supervised learning to directly optimize for control performance -- instead of minimizing the classification error -- and to avoid the use of branch-and-bound in both the offline and online phases of the algorithm.

The main contributions of the paper with regard to the literature are:

\begin{itemize}
    \item We propose a novel integrated reinforcement learning and MPC framework for control of mixed-logical dynamical systems. During online operation, the reinforcement learning agent simplifies a mixed-integer linear program into a linear program by fixing the discrete decision variables.
    \item The Q-function is partitioned across the prediction horizon, and the definition of decoupled Q-functions is conceived to make the learning problem tractable. The decoupled Q-function is approximated by a recurrent neural network, and its role in the reinforcement learning algorithm is described.  
    \item Simulation experiments in a microgrid system show the efficacy of the proposed approach in reducing the computational load of the MPC controller. Moreover, the comparison between the proposed approach and a method based on supervised learning reveals a trade-off in the case study: while the former outperforms in terms of feasibility, the latter outperforms in terms of optimality.
\end{itemize}


\subsection{Outline}

This paper is organized as follows. In Section \ref{sec:related_work}, we give an overview of the literature on the intersection of learning and control for hybrid systems. Section \ref{sec:control_problem} formalizes the control problem. Our novel method that integrates reinforcement learning and MPC for control of mixed-logical dynamical systems is described in Section \ref{sec:method}. Section \ref{sec:microgrid} presents the simulation setup, the results, and the discussion. This paper ends in Section \ref{sec:conclusions} with conclusions and suggestions for future work.

\section{RELATED WORK}\label{sec:related_work}

Some works have applied learning to reduce the solution time of mixed-integer programs by embedding learning into the branch-and-bound algorithm to substitute expert-designed heuristics with learned rules. For instance, learning can be used to improve cutting plane rules \cite{tang20a}, branching variable selection \cite{zhang2023survey}, and node selection \cite{balcan18learning2branch}. 
Although the aforementioned approaches indeed reduce the solution time of the considered mixed-integer programs, control applications often require a more expressive reduction in the computational load.

In \cite{menta2021} the state-action value function (Q-function) of time-invariant mixed-logical dynamics systems is approximated. The core of their approach is to parametrize the Q-function with Benders cuts and estimate the Q-function from a lower bound. However, the scope of application of this approach is limited because the paper does not consider discrete states and inputs. In \cite{Russo2023LAMPOs}, the computational burden of MPC is reduced by learning a state-dependent horizon and a state-dependent recomputation policy, i.e., the policy that decides whether the MPC optimal control problem should be recomputed at a given time step. An MPC controller is used as a function approximator of the state-action value in \cite{gros2022learning}, where the parameters of the controller are tuned by policy gradient methods. An extension of the same approach to mixed-integer problems was made in \cite{groszanon20_MixedIntegerRL}; however, the goal is to target performance rather than reduction in the computational cost of the optimization problem.

There is also a body of work that explores the use of reinforcement learning to jointly learn discrete and continuous policies \cite{masson2016reinforcement, fan2019hybridRL, neunert2020continuous}. In the artificial intelligence community, this problem is explored under the framework of Markov decision processes with parameterized actions. The problem is very similar to that of control of hybrid systems due to the nature of the action space, which has discrete and continuous elements. Even though these works can offer useful insight into the parametrization and training of policies for hybrid systems, they lack the optimality and constraint satisfaction that MPC can provide.

Outside the domain of hybrid systems, other works have also integrated learning into the MPC framework in various ways. The most popular approaches include the adaptation of the system model, the use of MPC as a safety filter for RL, and the online tuning of the cost and constraint functions for performance, see \cite{hewing2020, gorges2017} for more methods on the interplay of learning and MPC. These approaches are aimed towards improving performance, safety, and/or robustness and do not address the computational issues, which is the main concern of our work.

In the literature regarding energy systems, several approaches exist for the integration of learning and optimization-based control. In \cite{cai2023learning}, a parameterized MPC controller is tuned by an actor-critic reinforcement learning method to improve the performance of a home energy management system. A similar approach is developed in \cite{cai2023energy} for a residential microgrid system. Therein, in addition to the parameterized MPC approach, the energy management problem is framed as a cooperative coalition game, and the Shapley value is used to distribute the costs between the consumers. The authors of \cite{shengren2023optimal} propose the MIP-DQL algorithm in which the Q-function maximization problem is represented as a mixed-integer program, in which the system constraints are embedded. In this fashion, the reinforcement learning algorithm can guarantee operational constraints via online optimization. Another use of learning is to estimate prediction models for the MPC controller. In \cite{talib2023grey}, grey-box and neural network models are compared for multistep-ahead prediction for control of heating, ventilation, and air-conditioning systems. The paper \cite{huo2023learning} proposes the use of learning to assist column generation in the solution of the mixed-integer programs with the aim of reducing the computational cost of MPC-based approaches in energy management for microgrids. 

The use of mixed-logical dynamical systems for modeling and control of microgrids has been considered in \cite{parisio2014model}, \cite{pippia2019RuleBased}, 
\cite{masti2020}, and \cite{masti19_warmStart}. In these works, an MLD system is employed to cast a mixed-integer linear problem for microgrid operation optimization. A similar model is used in the case study of our work to solve the optimal dispatch problem.

\section{CONTROL PROBLEM}\label{sec:control_problem}

We consider MPC for its capacity to handle multivariable constrained hybrid systems.
In this context, mixed-logical dynamical (MLD) systems \cite{99bemporad} are typically used to formulate open-loop finite-horizon optimal control problems \cite{borrelli2017}.
Moreover, the equivalence of MLD systems and other hybrid system modeling frameworks was established in \cite{heemels01}, showing their broad applicability.
When MLD systems are used to formulate the MPC problem and the cost function and the constraints are linear, the resulting optimization problem is a mixed-integer linear program (MILP).
In this section, a general description of the MPC optimization problem for an MLD system is first addressed to set the stage for the formulation of the control problem as an MILP.

Consider the MLD system
\begin{equation}\label{eq:MLDsystem}
    \begin{split}        
        x(k+1) &= Ax(k) + B_1 u(k) + B_2 \delta (k) + B_3 z(k) + B_5, \\
        &E_2 \delta(k) + E_3 z(k) \leq E_1 u(k) + E_4 x(k) + E_5
    \end{split}
\end{equation}
where $x \in \mathbb{R}^{n_\mathrm{c}} \times \{0,1\}^{n_\mathrm{d}}$ is a vector containing the continuous and discrete system states, $u \in \mathbb{R}^{m_\mathrm{c}} \times \{0,1\}^{m_\mathrm{d}}$ are the continuous and discrete inputs, $\delta \in \{0,1\}^{r_\mathrm{d}}$ is a vector with the auxiliary discrete variables, $z \in \mathbb{R}^{r_\mathrm{c}}$ is a vector with the continuous auxiliary variables arising from the MLD modeling, and $A,\ \{B_{i}\}_{i=1,2,3,5},\ \{E_i\}_{i=1,2,3,4,5}$ are matrices of appropriate dimensions. Note that the linear constraints may represent logical constraints, a byproduct of MLD modeling, or system operating constraints. For a demonstration of the application of the discrete variables and auxiliary variables to represent logical constraints, the reader is referred to Appendix \ref{sec:sys_description}, where the model for the case study is characterized.

The goal is to control the system \eqref{eq:MLDsystem} with a receding-horizon strategy, where at each time step a finite-horizon optimal control problem is solved. Such an optimization problem can be formulated as follows:
\begin{equation}\label{eq:MPCgeneralproblem}
    \begin{split}
        \min_{\mathbf{x} (k), \epsc (k), \epsd (k)} \ 
        &J(\mathbf{x} (k), \epsc (k), \epsd (k))  \\
        \text{s.t. } & x(k+l+1) = Ax(k+l) + B_1 u(k+l) + \\
        &\hspace{1cm}+ B_2 \delta (k+l) + B_3 z (k+l) + B_5, \\
        &E_2 \delta(k+l) + E_3 z (k+l) \leq E_1 u (k+l) + \\ 
        & \hspace{3cm} + E_4 x (k+l) + E_5, \\
        & \text{for } l = 0,\dots,N_\mathrm{p} - 1,
    \end{split}
\end{equation}
where the cost function is defined by
\begin{equation}\label{eq:general_cost_function}
\begin{split}
    &J(\mathbf{x}(k), \epsc (k), \epsd (k)) =  \\ =&\sum_{l=0}^{N_\mathrm{p}-1} \left( \ell(x (k+l), u (k+l), \delta (k+l), z (k+l))\right) + \\ 
    & \hspace{3cm} + V_\mathrm{f}(x (k+N_\mathrm{p})),
\end{split}    
\end{equation}
where the prediction horizon is denoted by $N_\mathrm{p}$, the variable $l$ indexes the time along the prediction horizon, and the final cost is represented by $V_\mathrm{f}(\cdot)$. The predicted state trajectory over the prediction horizon is denoted by $\mathbf{x} (k) = [ x^\T (k), \ldots, x^\T (k+N_\mathrm{p}) ]^\T$. Let the predicted input be explicitly divided into continuous and discrete components $u(k) = [u_\mathrm{c}^\T(k),\ u_\mathrm{d}^\T(k)]^\T$. The stacked continuous input and auxiliary variables over the prediction horizon are represented in vector $\epsc (k) = [u_\mathrm{c}^\T (k), z^\T (k), \ldots, u_\mathrm{c}^\T (k+N_\mathrm{p}-1),\ z^\T (k+N_\mathrm{p}-1)]^\T$. Similarly, the stacked discrete inputs and auxiliary variables over the prediction horizon are expressed by $\epsd (k) = [u_\mathrm{d}^\T (k), \delta^\T (k), \ldots, u_\mathrm{d}^\T (k+N_\mathrm{p}-1),\ \delta^\T (k+N_\mathrm{p}-1)]^\T$.
For the sake of simplicity, the constraints are assumed to be polyhedral, as in \eqref{eq:MLDsystem}.

At each time step, the optimization \eqref{eq:MPCgeneralproblem} is solved, and the first entry of the decision variables is applied to the system. Let $(\mathbf{x}^*(k),\ \epsc^* (k),\ \epsd^* (k))$ be the optimal solution of \eqref{eq:MPCgeneralproblem}. The MPC control law then is defined by
\begin{equation}
    u_\mathrm{MPC}(k) = [\epsilon_{\mathrm{c},0}^{*,T}(k) ,\ \epsilon_{\mathrm{d},0}^{*,T}(k)]^\T
\end{equation}%
where the first entries of the solution vector are expressed as $[\epsilon_{\mathrm{c},0}^{*,T}(k) ,\ \epsilon_{\mathrm{d},0}^{*,T}(k)]^\T = [u_\mathrm{c}^{*,T} (k),\ z^{*,T} (k),\ u_\mathrm{d}^{*,T} (k),\ \delta^{*,T} (k)]^\T$.

If the cost function \eqref{eq:general_cost_function} is linear, e.g., state and terminal costs based on the 1-norm or the $\infty$-norm, then the optimization problem \eqref{eq:MPCgeneralproblem} can be conveniently recast as a mixed integer linear program \cite{borrelli2017}. The corresponding linear optimization problem in a compact form can be stated as follows:
\begin{equation}\label{eq:milp}
        \begin{split}
        \min_{\epsilon(k)} \ & c^\T\epsilon(k) \\
        \text{s.t. } & G\epsilon(k) \leq w + Sx(k)
        \end{split}
\end{equation}
where $\epsilon (k) = [\epsc ^\T(k),\ \epsd ^\T(k)]^\T$, and $c, \ G,\ w,\ S$ are matrices of appropriate sizes.

Note that the optimization problem \eqref{eq:milp} is parameterized by the state $x(k)$. Besides, the system state has an effect on the objective through the auxiliary variable $z(k)$, which is part of the decision variable $
\epsilon(k)$. Similarly, the resulting optimization problem can also possibly depend on an exogenous variable $\gamma (k)$, i.e., a measured external variable outside the influence of the controller. The effect of $\gamma(k)$ on the optimization problem is to further parameterize the mixed-integer program. In principle, the exogenous variable can potentially change any of the matrices $c,\ G, \ w,$ and $S$ that define \eqref{eq:milp}. In our problem formulation, we restrict the parametrization to the cost vectors $c$ and the constraint matrix $w$.

The problem \eqref{eq:milp} can then be extended to reflect the addition of the exogenous variable: 
\begin{equation}\label{eq:milp2}
        \begin{split}
        \min_{\epsilon(k)} \ & c^\T(
        \gamma(k))\epsilon(k) \\
        \text{s.t. } & G\epsilon(k) \leq w(\gamma(k)) + Sx(k)
        \end{split}
\end{equation}

For ease of notation, we leave out the explicit dependence of such parameters on the exogenous signal hereafter. Since \eqref{eq:milp2} is parameterized both by the system state $x(k)$ and the exogenous signal $\gamma (k)$, we define the augmented state
\begin{equation*}
	\chi(k) = [x^\T(k),\ \gamma^\T(k)]^\T
\end{equation*}
to fully characterize the parameters that define the parametric mixed-integer linear program.

The exogenous variables are necessary for the modeling of some applications, as shown in the case study in Section \ref{sec:microgrid}, where such variables are comprised of electricity prices, renewable energy generation, and energy demand forecasts.

\begin{remark}
    The assumption on the linearity of the objective function does not play any fundamental role in the development of the proposed approach in the next section. The choice for a linear objective is to simplify the notation and the presentation of this work. In principle, even a nonconvex objective cost could be chosen. Such a choice would, however, result in more complex optimization problems to be solved offline for learning and solved online for inference. 
    Similarly, even though \eqref{eq:milp2} can be formulated in our proposed approach with a more general parametrization in the exogenous variable $\gamma(k)$, we restrict the parametrization for the sake of simplicity and alignment with the case study.
\end{remark}

\section{METHOD: NOVEL INTEGRATION OF REINFORCEMENT LEARNING AND MPC}\label{sec:method}

In this section, we present the main contribution of our work: an integrated reinforcement learning and MPC method for the solution of mixed-integer linear problems for control of mixed-logical dynamical systems.
The main goal is to ease the online computational burden of solving mixed-integer linear programs by decoupling the computation of the integer and continuous decision variables. 
In our framework, the discrete decision variables are determined by reinforcement learning, and the continuous decision variables by MPC.
Accordingly, the solution of mixed-integer programs is entirely avoided in online operation. Consequently, our approach can be seen as an alternative to branch-and-bound for MPC for MLD systems.
In what follows, the decoupling of the discrete and continuous variables of the MILP of \eqref{eq:milp2} is addressed. Next, we explain the role of reinforcement in our approach.

Herein, it is assumed that the system dynamics are known and training takes place fully offline, i.e., there is no online adaptation of the learning parameters.

\subsection{Decoupling of decision variables}

The mixed-integer linear program (MILP) \eqref{eq:milp2} is parametric with respect to the initial state $x(k)$ and possibly an exogenous signal $\gamma(k)$.
Consider the set of these values in which \eqref{eq:milp2} is feasible. In this set, the solution of \eqref{eq:milp2} maps the signal $(x(k),\ \gamma (k))$ to their corresponding optimal decision variables $(\epsc (k)^*,\ \epsd (k)^*)$.

Consider the scenario where the discrete part of the optimal solution $\epsd (k)^*$ is known. Then the continuous part of the solution $\epsc (k)^*$ can be easily recovered by solving the problem \eqref{eq:milp2}, which becomes a linear program (LP) with the fixed discrete optimization variables, as shown below:
\begin{equation}\label{eq:lp}
	\begin{split}
		V(x(k),\ \gamma(k),\  &\epsilon_\mathrm{d}(k)^*) = \min_{\epsc(k)} \ c(\gamma(k))^\T[\epsc^\T(k),\ \epsd ^\T(k)^*]^\T \\
		\text{s.t. } & G[\epsc^\T(k),\ \epsd ^\T(k)^*]^\T \leq w(\gamma(k)) + S x(k)
	\end{split}
\end{equation}
In this manner, the optimal continuous decision variables can be found, provided that the optimal discrete decision variables are known. Evidently, assuming the knowledge of the discrete decision variables is not realistic. Nevertheless, we make use of this decoupling procedure by using learning approaches to provide approximate solutions to the discrete decision variables and by letting the optimization determine the continuous variables, as in \eqref{eq:lp}. One of the main goals of this work is to show that, with sufficient training data, an effective approximator for the discrete decision variables can be trained and that solving \eqref{eq:lp} with such approximated discrete variables yields solutions for the continuous variables without much sacrifice of optimality.

In the proposed approach, reinforcement learning is used to learn a policy that maps the system state to the discrete decision variables. The training of RL policy is done offline by repeated interaction with a model of the system and historical data. During online operation, the original MILP problem \eqref{eq:milp2} is turned into an LP problem \eqref{eq:lp} after the RL policy provides an approximate solution to the discrete decision variables given the current measured system state and exogenous variable. A more detailed description of the role of RL is given in the next subsection.

\begin{figure*}[htb]
    \centering
    \includegraphics[width=0.8\textwidth]{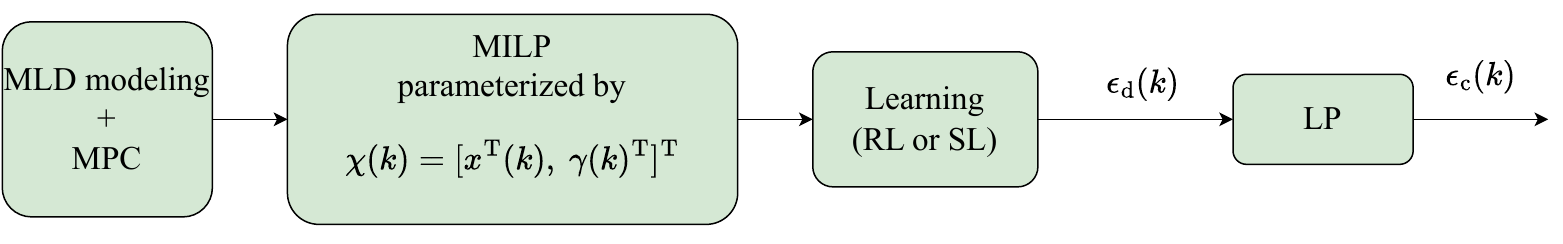}
    \caption{Representation of the decoupling of the discrete and continuous decision variables. From mixed logical dynamical (MLD) modeling and the use of an MPC approach for control, a mixed-integer linear program (MILP) can be formulated for the operation of the microgrid. Then, a learning approach—either reinforcement learning (RL) or supervised learning (SL) -- is used to determine the discrete variable $\epsilon_\mathrm{d}(k)$. The MILP is then simplified into a linear program (LP), which computes the continuous variable $\epsilon_\mathrm{c}(k)$. }
    \label{fig:milp2lp}
\end{figure*}

The main difficulty of solving MILPs is the discrete optimization variables, and modern solvers mostly rely on branch-and-bound as a strategy to address this problem. Therefore, computing these complicating decision variables with a learning-based method removes the main optimization hurdle. Furthermore, solving such LPs of the form \eqref{eq:lp} is computationally cheap with state-of-the-art specific-purpose solvers. The immediate benefit of the continuous variables being determined by an optimization-based control, such as MPC, is constraint satisfaction and optimality relative to the approximate discrete variables solution and subject to the feasibility of \eqref{eq:lp}. A depiction of the decoupling process is shown in Fig. \ref{fig:milp2lp}.

\subsection{Role of reinforcement learning}

Reinforcement learning (RL) is a general learning framework where the agent learns a control policy based on its interaction with the environment. It has received increasing attention in control applications due to its capacity to learn complex policies and for its low demand for online computation \cite{sutton2020}. Here we describe an approach that exploits the benefits of RL to efficiently determine the discrete decision variables of \eqref{eq:milp2}.

In the mixed-integer linear program \eqref{eq:milp2}, the main computational complexity stems from the number of discrete decision variables, which can come from both the number of actions per time step and the length of the prediction horizon.
In our method, reinforcement learning is used to ease the online computational burden of solving mixed-integer linear programs by decoupling the computation of the discrete and continuous decision variables. The discrete decision variables are determined by reinforcement learning, and the continuous decision variables by optimization-based control -- MPC. Accordingly, the solution of mixed-integer programs is entirely avoided in online operation.

\begin{figure*}[htb]
    \centering
    \includegraphics[width=0.7\textwidth]{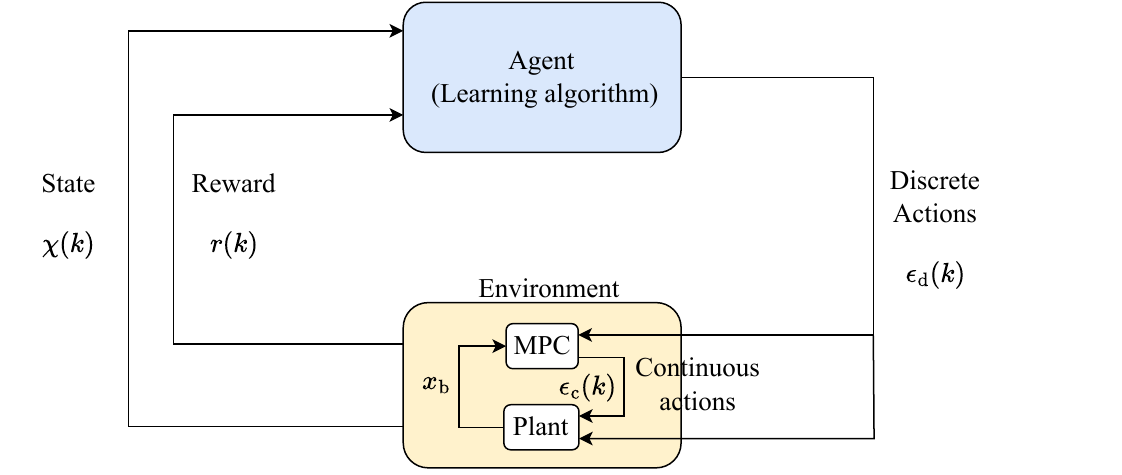}
    \caption{A depiction of the proposed control scheme that integrates reinforcement learning into an MPC framework. The agent's goal is to maximize its long-term reward. It learns to adapt its policy by repeatedly interacting with the environment, that is, by sending a discrete action $\epsilon_\mathrm{d}(k)$ and by receiving the extended state $\chi$ and immediate reward $r$. The MPC controller, which is lumped in the environment, receives this discrete action $\epsilon_\mathrm{d}(k)$ and then solves an optimization problem to determine the continuous action $\epsilon_\mathrm{c}(k)$. Finally, the input $\epsilon$ is fed to the system, and the next state is computed. 
    }
    \label{fig:RLtraining}
\end{figure*}

In order to formulate the problem as a Markov decision process (MDP) -- the standard reinforcement learning framework -- we lump together the system and the MPC controller in a single block to form the environment, see Fig. \ref{fig:RLtraining}. The environment receives the discrete actions $\epsd (k)$ from the RL agent and outputs the next state $\chi(k+1)$ and the corresponding reward $r(k)$. 
This abstraction allows the decoupling of the discrete and continuous decision variables, and it is crucial to bridge reinforcement learning and optimization-based control for MLD systems.
The agent is responsible for determining the discrete sequence of actions $\epsd (k)$ over the prediction horizon. This vector is then sent to the environment, which solves the optimization problem \eqref{eq:lp} and computes the continuous decision variables $\epsc (k)$. As a result, the environment outputs the next state $\chi(k+1)$ and the reward $r(k)$. Any reinforcement learning algorithm can be employed to train the agent; however, for simplicity, we use Deep Q-Learning \cite{mnih2013vanillaDQN} to present our approach. For a graphical representation of the reinforcement learning setting, see Fig. \ref{fig:RLtraining}.

Herein, the Q-function is defined as the expected reward over a finite horizon if the agent takes action $\epsd (k)$ at the current time step $k$ and then follows the policy $\epsd (k+l) = \pi(\epsd \ | \ \chi(k+l))$ over the remaining steps of the horizon:
\begin{equation}\label{eq:q_value}
\begin{split}
    Q^\pi (\chi,\ &\epsd) = \ \mathbb{E}_\pi  [\sum_{l=0}^{N_\mathrm{p}-1} \alpha^{l} r(k+l) \ | \ \chi(k) = \chi, \\ &\epsd(k)=\epsd, \ \epsd(k+l)=\pi(\epsd \ | \ \chi(k+l)) \ \forall l\neq 0  ]
\end{split}
\end{equation}
where $\alpha$ is the discount factor and the reward is defined as a function of the objective of \eqref{eq:lp}:
\begin{equation*}
    r(k) = f_\mathrm{reward}(\gamma(k), \epsilon(k))
\end{equation*}
where $\gamma(k)$ and $\epsilon(k)$ are respectively the exogenous variables and the optimization variables defined in the previous section, see \eqref{eq:milp2}. The system state $x(k)$ also indirectly influences the reward by modifying the constraints in \eqref{eq:lp} and consequently affecting the computation of the decision variables $\epsilon(k)$.
Moreover, $f_\mathrm{reward}(\cdot)$ is a scaling function used to keep the reward within reasonable ranges, e.g., $r \in [0,\ 1]$, preventing the gradients from becoming large during training, which would impair learning. If the action $\epsd (k)$ causes the LP of \eqref{eq:lp} to be infeasible, then the reward becomes negative, e.g., $r=-1$, to penalize this behavior. 
To approximate the undiscounted control problem, i.e., to appropriately consider long-term rewards, the discount factor $\gamma$ is typically chosen close to the upper limit of the interval $[0,1)$ \cite{sutton2020}.

The vector $\epsd (k)$ contains the entire sequence of discrete variables over the prediction horizon. 
For each time step of the prediction horizon $l=0, \ldots, N_\mathrm{p}-1$, we can represent the discrete sub-action per time step with $\varepsilon_{\mathrm{d},l}(k)$, hence
\begin{equation*}
    \epsd (k) = [\varepsilon_{\mathrm{d},0}^\T(k),\ \varepsilon_{\mathrm{d},1}^\T(k),\ \ldots,\ \varepsilon_{\mathrm{d},N_\mathrm{p}-1}^\T(k)]^\T \ .   
\end{equation*}
The potentially large size a combinatorial action space is a very challenging problem for RL algorithms. For instance, estimating the Q-function via neural networks, as it is typically implemented in value-based and actor-critic methods, would require a large and intractable number of units -- equal to the number of actions -- in the output layer.
Instead of using an RL algorithm to find the Q-function for the action $\epsd (k)$, as defined in \eqref{eq:q_value}, 
our approach shifts the goal to learning the decoupled Q-functions for the sub-actions $\{\varepsilon_{\mathrm{d},l}(k)\}_{l=0}^{N_\mathrm{p}-1}$.
Formally, the decoupled Q-function under a policy $\pi(\epsd \ | \ \chi(k))$ for a given time step $l \in \{0,...,N_\mathrm{p}-1\}$ of the prediction horizon is defined as
\begin{equation}\label{eq:decoupled_Q_function}
    \begin{split}
       Q_l^\pi(\chi, \ &\varepsilon) = \ \mathbb{E}_\pi [\sum_{i=0}^{N_\mathrm{p}-1} \alpha ^i r(i) \ | \ \chi(0) = \chi,  \ \varepsilon_{\mathrm{d},l}(0) = \varepsilon \ , \\ &
       \varepsilon_{\mathrm{d},j}(0) = [\pi(\epsd \ | \ \chi(i))]_j \ \  \forall j \neq l, \\ & \varepsilon_{\mathrm{d},j}(i) = [\pi(\epsd \ | \ \chi(i))]_j \ \ \forall j \land i \in \{1,...,N_\mathrm{p}-1\}, \\ &\epsc(i) \text{ is the solution of \eqref{eq:lp}},\ 
       \\ & \epsilon(i) = [\epsd^\T(i),\ \epsc^\T(i)]^\T,
       \\ &r(i) = f_\mathrm{reward}(\chi(i),\  \epsilon(i)),  \\ & \chi(i+1)=f_\mathrm{MDP}(\chi(i), \ \epsilon(i)) \ ]
    \end{split}     
\end{equation}
where $[\pi(\epsd \ | \ \chi(i))]_j$ is the $j$th element of the discrete decision variables vector. More specifically, the decoupled Q-function is defined as the expected return of sub-action $\varepsilon$ in state $\chi$ given that the first sub-action at index $l$ is $\varepsilon_{d,l}(0)=\varepsilon$, the initial state is $\chi(0)=\chi$, the other sub-actions $j\neq l$ are chosen under policy $\varepsilon_{\mathrm{d},j}(0) = [\pi(\epsd \ | \ \chi(k))]_j$ for the first time step ($k=0$) and that the system is guided by policy $\epsd(k) = \pi(\epsd \ | \ \chi(k))$ for $k \in \{1,...,N_\mathrm{p}-1\}$ under the MDP dynamics $\chi(k+1)= f_\mathrm{MDP}(\chi(k), \ \epsilon (k))$.
The decoupled Q-functions can then be used to determine the policies for each of the sub-actions, e.g., the greedy policies are defined as
\begin{equation}\label{eq:greedy_action_decoupled}
    \varepsilon_{\mathrm{d},l}(k) = \max_\varepsilon \ Q_l^\pi(\chi(k),\ \varepsilon) \ {\text{for } l=0,\ \ldots,\ N_\mathrm{p}-1}.
\end{equation}

\begin{figure*}[htb]
    \centering    \includegraphics[width=\textwidth]{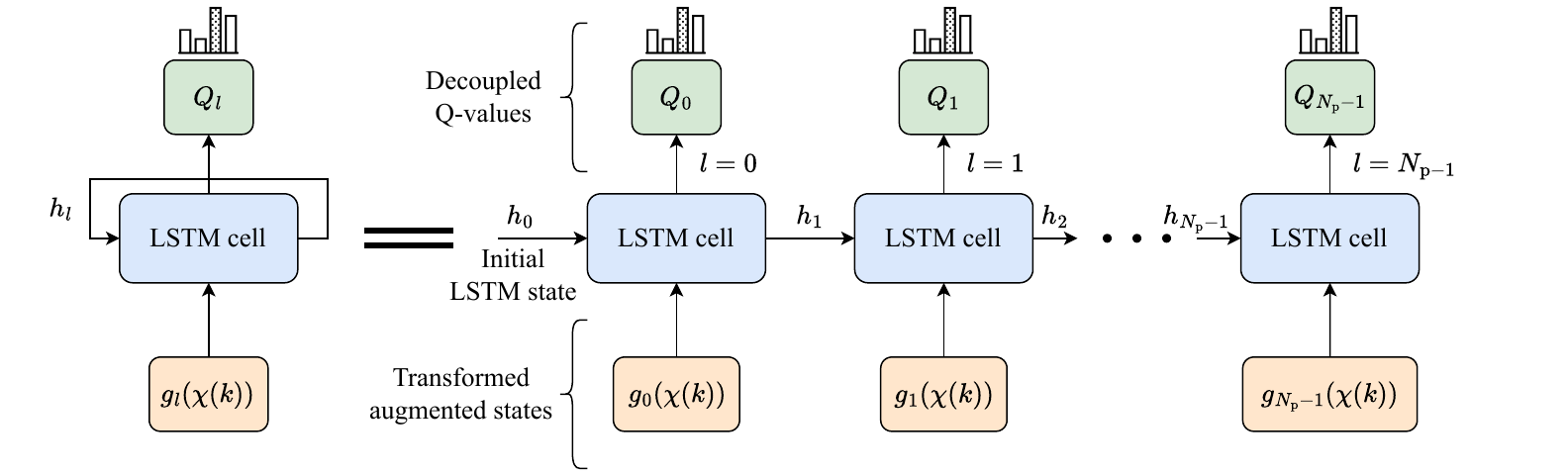}
    \caption{A representation of the recurrent LSTM network on the left-hand side and the unrolled LSTM network on the right-hand side. Note that the LSTM is unrolled for the duration of the prediction horizon $N_\mathrm{p}$. Moreover, at each time step $k$, the augmented state $\chi(k)$ can go through the operation $g_k(\cdot)$, changing the manner in which the augmented state is presented to the LSTM network at time step $k$. This can be interpreted as a preprocessing technique to better exploit the structure of our problem.}
    \label{fig:LSTM}
\end{figure*}

The definition \eqref{eq:decoupled_Q_function} is not tractable, but it can be useful for the design of a suitable approximator. The goal is to approximate each of the decoupled Q-functions $\{\tilde{Q}_l(\cdot)\}_{l=0}^{N_\mathrm{p}-1}$ over the prediction horizon.
By definition, the decoupled Q-function of time step $j$ depends on the sub-actions $l\in\{0,...,N_\mathrm{p}-1\}\setminus \{j\}$ taken over the prediction horizon via the policy $\pi(\cdot)$. Herein, we assume that the past sub-actions are more relevant to the current sub-action than the future sub-actions. Therefore, to compute the decoupled Q-function of the time step $j$, only the sub-actions up to time step $j$ of the prediction horizon $\{\varepsilon_{d,l}(k)\}_{l=0}^{j-1}$ are considered.
We propose the use of recurrent neural networks (RNNs) to recursively compute each of the decoupled Q-values for its capacity of carrying over information from the past through an internal state. In this fashion, the decoupled Q-function receives information from the past sub-actions.
Based on this assumption, an approximation of the decoupled Q-function with an RNN can be formulated as
\begin{equation}
    Q_l(\chi(k),\ \varepsilon_\mathrm{d,l} (k)) \approx \tilde{Q}_l(\chi(k),\ \varepsilon_\mathrm{d,l} (k),\ h_{l-1}(k))
\end{equation}
where the hidden state $h_{l-1}(k)$ condenses information about past sub-actions. The evaluation of each decoupled Q-function is illustrated in Fig. \ref{fig:LSTM}, which depicts the forward pass of the unrolled long short-term memory (LSTM) network, a type of RNN. In particular, the LSTM network is employed for its capacity of capturing long-term dependencies in data and its structure, which attenuates the vanishing gradient problem \cite{gers2000learning}. 
Notably, the online evaluation of this approximator involves only the forward pass of the LSTM network, ensuring computational efficiency and tractability.

While each step of the prediction horizon has its own decoupled Q-function $Q_l(\cdot)$, the signal $h_{l-1}(k)$ retains the dependence of the previous sub-actions. In this way, some degree of interdependence between the sub-actions is modeled, and the learning problem is simplified due to the decoupling of the Q-functions. Moreover, the integer and continuous decision variables are also not entirely decoupled since the training of the approximator is based on the augmented state and reward signals, which depend on all the decision variables. 
Therefore, in both cases, decoupling aids tractability while maintaining some degree of interconnection, which is arguably desirable in our setting.

The idea of using LSTM networks to decouple the actions of each time step of the prediction horizon was first explored in \cite{cauligi21_CoCo} in the context of supervised learning. In contrast, our approach employs the LSTM network as a function approximator of the decoupled Q-functions in the setting of reinforcement learning.

With the approximator described, the training procedure using a value-based RL method (Deep Q-Learning \cite{mnih2013vanillaDQN}) is discussed. Each episode of the training procedure starts with a randomly chosen initial state $\chi(k)$.
Then, the RL agent repeatedly proposes a discrete action $\epsilon_\mathrm{d}(k)$, sends it to the environment, and collects the reward $r(k)$ and the next state $\chi(k+1)$. An episode ends when a pre-determined time limit is reached or when the RL agent suggests a discrete action that makes the optimization problem \eqref{eq:lp} infeasible. At each time step, the transitions $(\chi(k),\ \epsilon_{\mathrm{d}}(k),\ r(k),\ \chi(k+1))$ are stored in a data buffer $\mathcal{D}$, which only holds a finite number of the most recent transitions. After the end of each episode, $N_\mathrm{batch}$ transitions are randomly sampled from $\mathcal{D}$ and the weights of the LSTM network are updated by gradient descent on the following loss function:
\begin{equation*}
	L = \frac{1}{N_\mathrm{batch}} \sum_{k=1}^{N_\mathrm{batch}}  \sum_{l=0}^{N_\mathrm{p}-1} (y_l(k) - \tilde{Q}_l(\chi(k),\ \varepsilon_{\mathrm{d},l}(k)) )^2
\end{equation*}
where the temporal difference (TD) target is defined as
\begin{equation*}
	y_l(k) = r(k) + \alpha  \max_\varepsilon \tilde{Q}_l(\chi(k+1),\ \varepsilon)
\end{equation*} 
and from now on, the dependency of the decoupled Q-functions on the hidden state $h$ is omitted for ease of notation. 
Some alternative definitions of the TD target can be found in \cite{tavakoli2018action}. The training of the RL agent is deemed complete when a predetermined number of episodes or number of transitions is reached.

Once training is completed, the RL agent is employed to suggest the discrete action $\epsd (k)$ and, consequently, the MPC optimization problem \eqref{eq:MPCgeneralproblem} is simplified to the LP described in \eqref{eq:lp}. As a result, the solution time of the MILP is replaced by the time of inference of an LSTM network and the solution time of an LP, which can be efficiently solved by modern solvers, greatly reducing the computation time.

The offline training and the online inference are described in Algorithms \ref{alg:offline} and \ref{alg:online}, respectively.

\begin{algorithm}[htb]
    \caption{Offline Training}
    \label{alg:offline}
    \begin{algorithmic}[1]
        \State Initialize recurrent neural network that estimates the decoupled Q-function $\tilde{Q}(\chi,\ \epsilon;\ \theta)$ with random weights
        \State Initialize databuffer $\mathcal{D} = \{\}$ 
        \For{episode = $1, ...,N_\mathrm{episodes}$}
            \State Set random state $\chi(0)$ \For{$k=1,....,N_\mathrm{steps} $}
                    \For{$l$ from $0$ to $N_\mathrm{p}-1$}
                    \State $\varepsilon_{\mathrm{d},l}(k) \sim \mathrm{softmax} \ \tilde{Q}_l(\chi(k), \ \epsilon_{\mathrm{d},l}; \ \theta)$
                \EndFor
                \State Set $\epsd(k) = [\varepsilon_{\mathrm{d},0}^\T(k),\ \varepsilon_{\mathrm{d},1}^\T(k),\ \ldots,\ \varepsilon_{\mathrm{d},N_\mathrm{p}-1}^\T(k)]^\T$
                \State Get $\epsc(k)$ by solving the linear program \eqref{eq:lp}
                \State Set $\epsilon(k) = [\epsd^\T(k),\ \epsc^\T(k)]^\T$
                \State Get reward $r(k) = f_\mathrm{reward}(\chi(k),\ \epsilon(k))$
                \State Get next state $\chi(k+1) = f_\mathrm{MDP}(\chi(k),\ \epsilon(k))$
                \State Store $(\chi(k), \epsd(k), r(k), \chi(k+1))$ in $\mathcal{D}$
                \State Sample a minibatch of transitions $\{(\chi^j, \epsd^j, r^j, \chi^{j+1})\}_{j=1}^{N_\mathrm{batch}}$ from $\mathcal{D}$
                \State Set $y^j_l = 
                    r^j$ for terminal $\chi^{j+1}$ or $ y^j_l =
                    r^j_l + \alpha \max_\epsilon Q_l(\chi^{j+1}, \epsilon; \ \theta)$ for non-terminal $\chi^{j+1}$ for $l=0,...,N_\mathrm{p}-1$ and $j=1,...,N_\mathrm{batch}$
                \State Perform a gradient descent step on $\sum_{j=1}^{N_\mathrm{batch}}\sum_{l=0}^{N_\mathrm{p}-1} (y^j_l - \tilde{Q_l}(\chi^{j}, \varepsilon_{\mathrm{d},l}^j;\ \theta))^2$ to update the network weights $\theta$
            \EndFor
        \EndFor
    \end{algorithmic}
\end{algorithm}

\begin{algorithm}[htb]
    \caption{Online Inference}
    \label{alg:online}
    \begin{algorithmic}[1] 
            \State Measure state $\chi(k)$
            \For{$l$ from $0$ to $N_\mathrm{p}-1$} \Comment{\textit{Compute sub-actions}}
                \State $\varepsilon_{\mathrm{d},l}(k) = \arg\max_\varepsilon \ \tilde{Q}_l(\chi(k), \ \varepsilon)$
            \EndFor
            \State Set = $\epsd(k) = [\varepsilon_{\mathrm{d},0}^\T(k),\ \varepsilon_{\mathrm{d},1}^\T(k),\ \ldots,\ \varepsilon_{\mathrm{d},N_\mathrm{p}-1}^\T(k)]^\T$
            \State Get $\epsc(k)$ by solving the linear program \eqref{eq:lp}
            \State \textbf{return} $\epsd(k)$, $\epsc(k)$
    \end{algorithmic}
\end{algorithm}

\subsection{Differences between methods based on supervised and reinforcement learning}

To shed light on the fundamental differences between methods based on supervised learning (SL), such as \cite{cauligi22_PRISM}, and our approach, we explain the key differences in the learning setups.
From the perspective of supervised learning (SL), the problem is formulated as a classification task, where the network predicts the optimal sub-action at each time step within the prediction horizon. Rather than relying on decoupled Q-values to derive sub-actions, SL directly approximates the mapping from the augmented state to the corresponding optimal sub-actions. This approach requires repeatedly solving the MILP \eqref{eq:milp2} to optimality in order to construct the training dataset, which can be computationally expensive. Furthermore, unlike reinforcement learning, where the objective is to minimize the long-term operational cost of the system, SL focuses solely on minimizing the classification error during training. 
For a brief description of the training process in the context of supervised learning, the reader is referred to Appendix \ref{sec:supervised_learning}.

\section{CASE STUDY: MICROGRID CONTROL}\label{sec:microgrid}

In the case study, we aim to solve the economic power dispatch problem for a microgrid system, i.e., to fulfill the power demand of the microgrid loads while minimizing the operation cost. The system is described in Appendix \ref{sec:sys_description} and depicted in Fig. \ref{fig:microgrid}. The knowledge of the market energy prices and the forecasts of the power demand and power generation from renewable energy sources -- solar and wind energy are considered -- allows the operator to plan and optimally schedule the operation of the microgrid over a time horizon. To accomplish this goal, the microgrid operator has to determine a set of discrete and continuous actions over a finite prediction horizon. The discrete control actions are the following: buying or selling from grid $\delta_\mathrm{grid}(k)$, charging or discharging the energy storage system $\delta_\mathrm{b}(k)$ and turning on or turning off the generators $\{\delta_i^\mathrm{dis}(k)\}_{i=1}^{N_\mathrm{gen}}$ at time step $k$. The continuous actions are the power exchange with the main grid $P_\mathrm{grid}(k)$, power exchange with the energy storage system $P_\mathrm{b}(k)$, and the power provided by the dispatchable generators $P_i^\mathrm{dis}(k)$. For a summary of the decision variables, see Table \ref{tab:decision_variables}. In this setting, the decision variables are represented by the vectors $\epsc (k)$ and $\epsd (k)$, which are formed by the concatenation of all discrete and continuous decision variables, respectively, over the prediction horizon.
Furthermore, the exogenous signal $\gamma(k)$ represents the forecasts over the prediction horizon for the: purchasing, sale, and production prices; power generation from the renewable energy sources; and power demand from the loads.

\begin{table}[htb]
    \centering
    \fontsize{10pt}{10pt}\selectfont
    \begin{tabular}{|c|c|}
        \hline
         Variable & Description \\
         \hline
         $\delta_\mathrm{b}$ & status of the storage unit: \\ & charging(1)/discharging(0)\\
         $\delta_\mathrm{grid}$ & mode of connection with main grid: \\
         &importing(1)/exporting(0) \\
         $\delta_i^{\mathrm{dis}}$ & state of dispatchable unit $i$: on(1)/off(0) \\
         $P_\mathrm{b}$ & power exchanged with the storage unit [kW] \\
         $P_\mathrm{grid}$ & power exchanged with the main grid [kW] \\
         $P_i^\mathrm{dis}$ & power generated by dispatchable unit $i$ [kW] \\
         \hline
    \end{tabular}
    \caption{Decision variables and their descriptions.}
    \label{tab:decision_variables}
\end{table}

We consider the receding horizon control strategy laid out in Section \ref{sec:control_problem}. Consequently, the control problem can be formulated as the MILP of \eqref{eq:milp2}. Efficiently solving such an optimization problem is at the core of the problem.
In this context, we compare our RL-based approach against standard MPC for MLD systems, where the optimization problem \eqref{eq:MPCgeneralproblem} is solved to optimality with branch-and-bound, and against the supervised learning (SL) proposed in \cite{cauligi22_PRISM} briefly described in the last section and in Appendix \ref{sec:supervised_learning}.

\begin{figure}[htb]
    \centering
    \includegraphics[width=0.3\textwidth]{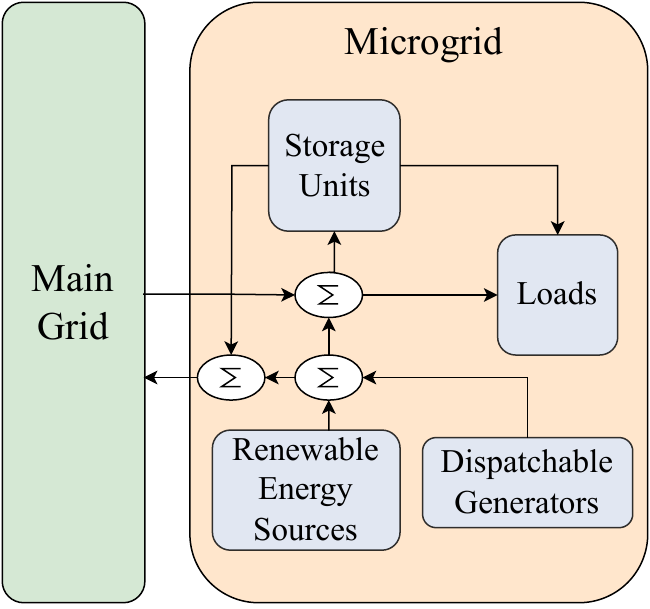}
    \caption{Depiction of the elements of a microgrid and a bidirectional connection with the main grid. The pointed arrows indicate the possibility of power flow between two elements.}
    \label{fig:microgrid}
\end{figure}

\subsection{Setup}

The renewable energy generation and load profiles were taken from the actual operation of the Dutch main grid, which is publicly available at \cite{entso}.
Since the microgrid considered in this case study does not have the capacity to meet the demand of the entire grid, the real power profiles were linearly downscaled to adjust the values to our case study.
This is the only preprocessing operation applied to these signals, so the shapes of the renewable energy generation and load profiles remain unchanged.
The RL-based and SL-based algorithms are trained on one year of data (2022) and their performances are evaluated in a different year (2021) to assess the capacity of the approaches to generalize and adapt to new data.

The price profiles -- $c_\mathrm{prod},\ c_\mathrm{buy},\ c_\mathrm{sell}$ -- are not publicly available, and, consequently, they were synthetically generated by
three different normal distributions that obey a basic principle: on average, producing energy is cheaper than purchasing energy from the main grid and higher than selling energy to the main grid -- examples of the cost profiles can be seen in Fig. \ref{fig:priceprofiles} in Appendix \ref{sec:sys_description}.
Due to the randomness of the prices, which mimic real market conditions, the price order can change, and the microgrid operator can take advantage of some occasions. 
For instance, when the production price is smaller than the sell price, it is profitable to generate power that exceeds the load demand and sell the power surplus to the main grid. The dispatchable units are assumed to be identical, and, consequently, the same production price is used for all of them.

Furthermore, the microgrid model, which is described in Appendix \ref{sec:sys_description}, is used for the training of the supervised learning and reinforcement learning agents, for the MPC controller, and for the system simulation. 
This ensures that performance differences arise solely from the algorithms. 
For the RL-based approach, Boltzmann exploration is used during training so that exploration occurs according to the decoupled Q-values. In this type of exploration, the probability of a given sub-action being chosen is proportional to the softmax of its decoupled Q-value. The policy used for exploration is described below
\begin{equation*}
	\pi^{\mathrm{RL}} (\varepsilon_{\mathrm{d},l,k}(k) \ | \ \chi(k) ) \propto e^{\xi \cdot {Q_l(\chi(k),\ \varepsilon_{\mathrm{d},l,k}(k))}}
\end{equation*}
where $\pi^{\mathrm{RL}}(\varepsilon_{\mathrm{d},l,k}(k) \ | \ \chi(k))$ represents the probability of selection of the sub-action $\varepsilon_{\mathrm{d},l,k}(k)$ in state $\chi(k)$, and $\xi$ is the exploration temperature. At the beginning of training, the temperature is chosen to $\xi = 0$ so that the probabilities of selecting the sub-actions are equal, which is equivalent to random exploration. As training progresses, the temperature is gradually raised to incentivize greedier behavior.

For the SL and RL approaches, the neural network architecture consists of a single layer of an LSTM network, whose outputs are connected to a fully connected layer. The approaches are then trained with several sizes of the hidden state of the LSTM network in the set $\{64,128,256,512\}$. To simplify the presentation of the results, we only report the best-performing trained neural networks after hyperparameter tuning.

The methods are simulated with four prediction horizons $N_\mathrm{p} \in \{4,12,24,48\}$, corresponding to look-ahead periods of $2,\ 6,\ 12$, and $24$ hours, respectively, given the sampling time of $T_\mathrm{s} = 30$ minutes. The aim of these experiments is to assess how the effectiveness of the methods scales with the increase of the number of binary variables. 

To ensure fair comparison, we use the general-purpose solver GUROBI \cite{gurobi} for the corresponding LPs and MILPs. 

\subsection{Simulation results}

The simulation results for the methods and different prediction horizons are shown in Table \ref{tab:results} and Figs. \ref{fig:opt_gap}, \ref{fig:infeas_rate}, and \ref{fig:computation_time}. As performance metrics, we analyze the optimality gap, infeasibility rate, and computation time. The optimality gap is the relative distance to the optimal solution computed by branch-and-bound, i.e.
\begin{equation*}
    \mathrm{optimality \ gap} = \frac{J_{\mathrm{learning}}-J_{\mathrm{optimal}}}{ J_{\mathrm{optimal}} } \times 100,
\end{equation*}
where $J_\mathrm{learning}$ is the cost of the linear program \eqref{eq:lp} when a learning method -- either RL or SL -- is used to determine the discrete variables and $J_{\mathrm{optimal}}$ is the cost of the mixed-integer program \eqref{eq:milp2} when it is solved to optimality with branch-and-bound.
The infeasibility rate is an empirical probability of the learning methods providing discrete sequences $\epsd (k)$ that result in an infeasible linear program \eqref{eq:lp}.
Computation time is also relative to the time necessary for the branch-and-bound solver to reach optimality, i.e.
\begin{equation*}
    \mathrm{reduction \ factor} = \frac{T^\mathrm{max}_\mathrm{optimal}}{T^\mathrm{max}_\mathrm{i}} \quad \text{for} \quad i\in \{\text{RL, SL}\},
\end{equation*}
where $T^\mathrm{max}_\mathrm{i}$ is the maximum of the inference time of the LSTM added to the solution time of the linear program \eqref{eq:lp} and $T^\mathrm{max}_\mathrm{optimal}$ is the maximum solution time of the mixed-integer program \eqref{eq:milp2}. 

The simulation experiments, conducted over $3000$ randomly selected initial states and summarized in Table \ref{tab:results} demonstrate the effectiveness of learning-based approaches in significantly reducing the online computation time of MPC. In control applications, often the maximum computation is a more important criterion than the average computation time for hardware specification. As shown in in Fig. \ref{fig:computation_time}, which reports the reduction in the maximum computation time, the learning-based approaches are $8$ to $16$ faster than the branch-and-bound solver in the worst-case.

\begin{table*}[htb]
\centering
\resizebox{0.65\textwidth}{!}{%
\begin{tabular}{cl|l|l|l}
\multicolumn{2}{l|}{} &
  \multicolumn{1}{c|}{\begin{tabular}[c]{@{}c@{}}Optimality gap\\  (\%)\end{tabular}} &
  \multicolumn{1}{c|}{\begin{tabular}[c]{@{}c@{}}Infeasibility rate \\ (1/1000)\end{tabular}} &
  \multicolumn{1}{c}{\begin{tabular}[c]{@{}c@{}}Computation time\\  (mean/max/std)\end{tabular}} \\ \hline
\multicolumn{1}{c|}{\multirow{3}{*}{$N_\mathrm{p}=4$}}  & RL      & 0.20 & 1.6   & 1.2 / 2.1 / 0.1     \\ \cline{2-5} 
\multicolumn{1}{c|}{}                                   & SL      & 0.04 & 24.6  & 1.3 / 2.3 / 0.1     \\ \cline{2-5} 
\multicolumn{1}{c|}{}                                   & Optimal & 0.00 & 0.0   & 3.4 / 19.7 / 0.7    \\ \hline
\multicolumn{1}{c|}{\multirow{3}{*}{$N_\mathrm{p}=12$}} & RL      & 0.36 & 2.3   & 2.0 / 3.6 / 0.2     \\ \cline{2-5} 
\multicolumn{1}{c|}{}                                   & SL      & 0.12 & 12.3  & 2.1 / 3.2 / 0.2     \\ \cline{2-5} 
\multicolumn{1}{c|}{}                                   & Optimal & 0.00 & 0.0   & 8.0 / 46.5 / 2.3    \\ \hline
\multicolumn{1}{c|}{\multirow{3}{*}{$N_\mathrm{p}=24$}} & RL      & 0.59 & 1.0   & 4.8 / 9.4 / 0.4     \\ \cline{2-5} 
\multicolumn{1}{c|}{}                                   & SL      & 0.15 & 54.3  & 4.3 / 10.8 / 0.4    \\ \cline{2-5} 
\multicolumn{1}{c|}{}                                   & Optimal & 0.00 & 0.0   & 23.6 / 84.8 / 3.3   \\ \hline
\multicolumn{1}{c|}{\multirow{3}{*}{$N_\mathrm{p}=48$}} & RL      & 0.76 & 11.0  & 10.9 / 14.5 / 0.5   \\ \cline{2-5} 
\multicolumn{1}{c|}{}                                   & SL      & 0.18 & 93.67 & 10.0 / 14.3 / 0.6   \\ \cline{2-5} 
\multicolumn{1}{c|}{}                                   & Optimal & 0.00 & 0.0   & 116.2 / 231.3 / 8.5
\end{tabular}%
}
\caption{Simulation results for the reinforcement learning (RL), supervised learning (SL), and branch-and-bound approaches for solving the mixed-integer linear program \eqref{eq:milp2}. The optimality gap is the relative distance between the objectives of the linear program \eqref{eq:lp} using the learning approaches and the objective of the mixed-integer program \eqref{eq:milp2} solved to optimality with branch-and-bound. The infeasibility rate is the frequency at which the learning approaches propose discrete actions that cause linear program \eqref{eq:lp} to be infeasible. 
The mean, maximum, and standard deviation of the computation time are reported.
}
\label{tab:results}
\end{table*}

\begin{figure}[htb]
\centering
\begin{tikzpicture}
\begin{axis}[
    width=\columnwidth,
    height=5cm,
    xlabel={Prediction horizon},
    ylabel={Optimality gap ($\%$)},
    grid=both,
    minor grid style={gray!25},
    major grid style={gray!50},
    legend pos=north west,
    legend style={font=\scriptsize},
    tick label style={font=\scriptsize},
    label style={font=\footnotesize},
    title style={font=\footnotesize},
    title={},
    xtick={1,2,3,4},
    xticklabels={$4$, $12$, $24$, $48$}
]

\addplot[mark=triangle*, mark size=3pt, blue, dotted]
coordinates {
    (1, 0.2)
    (2, 0.36)
    (3, 0.59)
    (4, 0.79)
};
\addlegendentry{RL};

\addplot[mark=diamond*, mark size=3pt, red, dotted]
coordinates {
    (1, 0.04)
    (2, 0.12)
    (3, 0.15)
    (4, 0.18)
};
\addlegendentry{SL};

\end{axis}
\end{tikzpicture}
\caption{Optimality gap for different prediction horizons for the reinforcement learning (RL) and supervised learning (SL) approaches in the case study.}
\label{fig:opt_gap}
\end{figure}
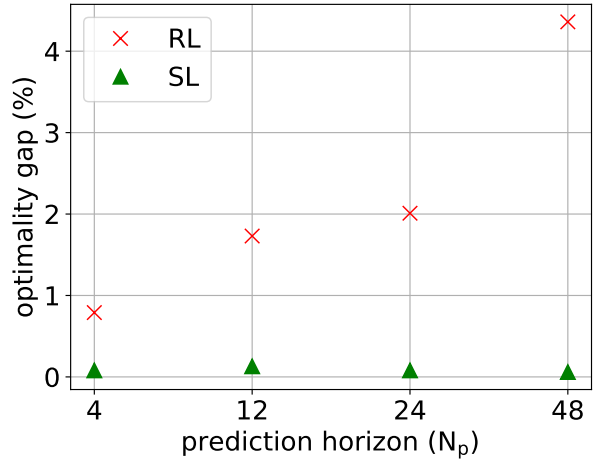

\begin{figure}[htb]
\centering
\begin{tikzpicture}
\begin{axis}[
    width=\columnwidth,
    height=5cm,
    xlabel={Prediction horizon},
    ylabel={Infeasibility rate ($\cdot/1000$)},
    grid=both,
    minor grid style={gray!25},
    major grid style={gray!50},
    legend pos=north west,
    legend style={font=\scriptsize},
    tick label style={font=\scriptsize},
    label style={font=\footnotesize},
    title style={font=\footnotesize},
    title={},
    xtick={1,2,3,4},
    xticklabels={$4$, $12$, $24$, $48$}
]

\addplot[mark=triangle*, mark size=3pt, blue, dotted]
coordinates {
    (1, 1.6)
    (2, 2.3)
    (3, 1.0)
    (4, 11.0)
};
\addlegendentry{RL};

\addplot[mark=diamond*, mark size=3pt, red, dotted]
coordinates {
    (1, 24.6)
    (2, 12.3)
    (3, 54.3)
    (4, 93.67)
};
\addlegendentry{SL};

\end{axis}
\end{tikzpicture}
\caption{Infeasibility rate for different prediction horizons for the reinforcement learning (RL) and supervised learning (SL) approaches in the case study.}
\label{fig:infeas_rate}
\end{figure}
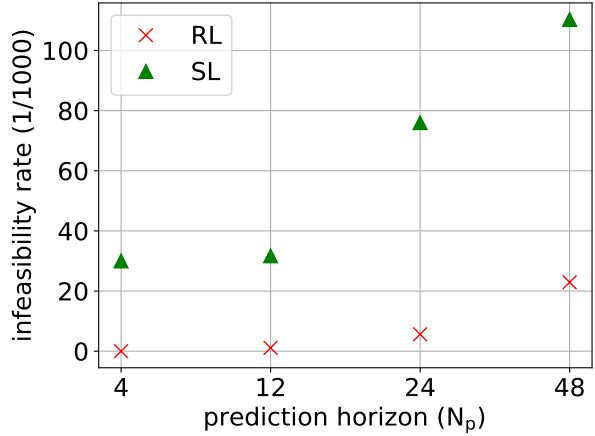

\begin{figure}[htb]
\centering
\begin{tikzpicture}
\begin{axis}[
    width=\columnwidth,
    height=5cm,
    xlabel={Prediction horizon},
    ylabel={Reduction factor},
    grid=both,
    minor grid style={gray!25},
    major grid style={gray!50},
    legend pos=north west,
    legend style={font=\scriptsize},
    tick label style={font=\scriptsize},
    label style={font=\footnotesize},
    title style={font=\footnotesize},
    title={Reduction in maximum computation time},
    xtick={1,2,3,4},
    xticklabels={$4$, $12$, $24$, $48$}
]

\addplot[mark=triangle*, mark size=3pt, blue, dotted]
coordinates {
    (1, 9.38)
    (2, 12.91)
    (3, 9.02)
    (4, 15.95)
};
\addlegendentry{RL};

\addplot[mark=diamond*, mark size=3pt, red, dotted]
coordinates {
    (1, 8.56)
    (2, 14.53)
    (3, 7.85)
    (4, 16.17)
};
\addlegendentry{SL};

\end{axis}
\end{tikzpicture}
\caption{Reduction factor of the maximum computation time $\frac{T^\mathrm{max}_\mathrm{optimal}}{T^\mathrm{max}_i}$, $i \in \{\text{RL, SL}\}$, for different prediction horizons for the reinforcement learning (RL) and supervised learning (SL) approaches in the case study.}
\label{fig:computation_time}
\end{figure}

\begin{figure}
    \centering
    \includegraphics[width=\linewidth]{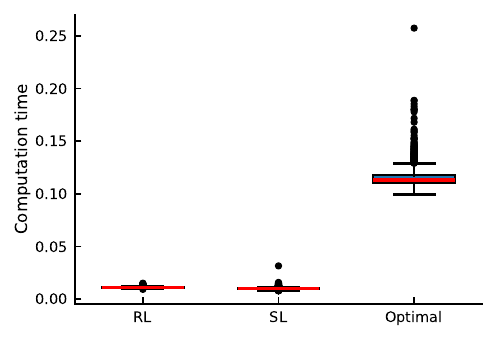}
    \caption{A box plot containing the computation time for the RL-based, SL-based, and MILP approaches. The median is highlighted in red. The blue box, only visible for the optimal (MILP) approach, extends from the first quartile to the fourth quartile, meaning that $50\%$ of the data lies in this region. The black points represent outliers. The black vertical lines represent the range of data, excluding outliers.}
    \label{fig:boxplot}
\end{figure}

\begin{figure}[htb]
    \centering
    \begin{subfigure}[b]{0.48\textwidth}
    \includegraphics[width=\textwidth]{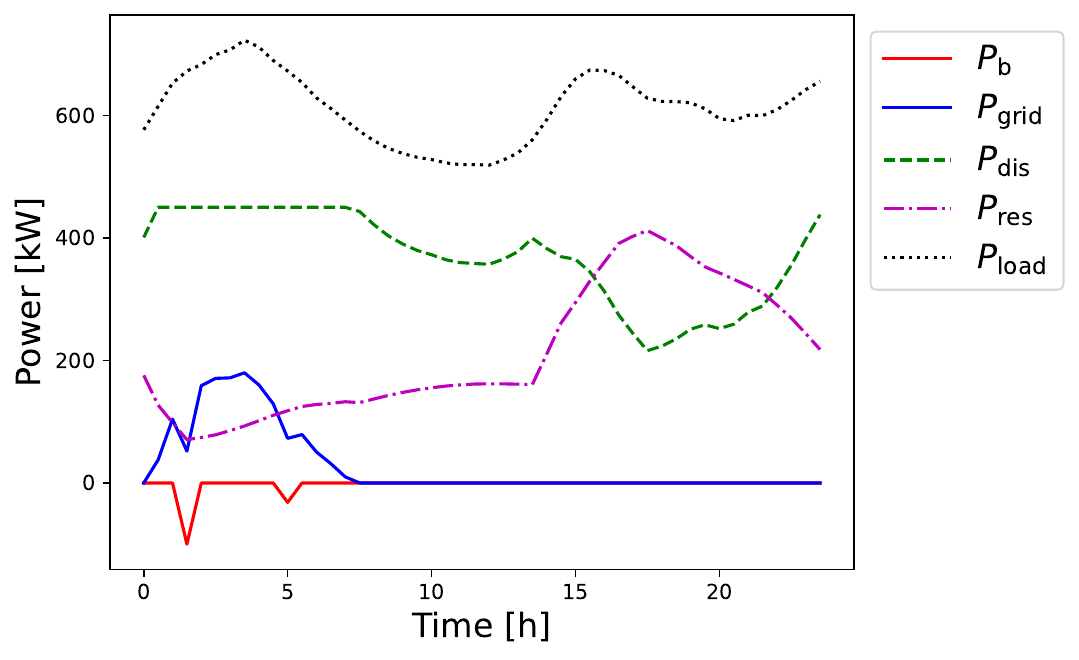}
    \caption{Optimal solution (MILP).}
    \label{subfig:milp}
    \end{subfigure}
    \hfill
    \begin{subfigure}[b]{0.48\textwidth}
    \includegraphics[width=\textwidth]{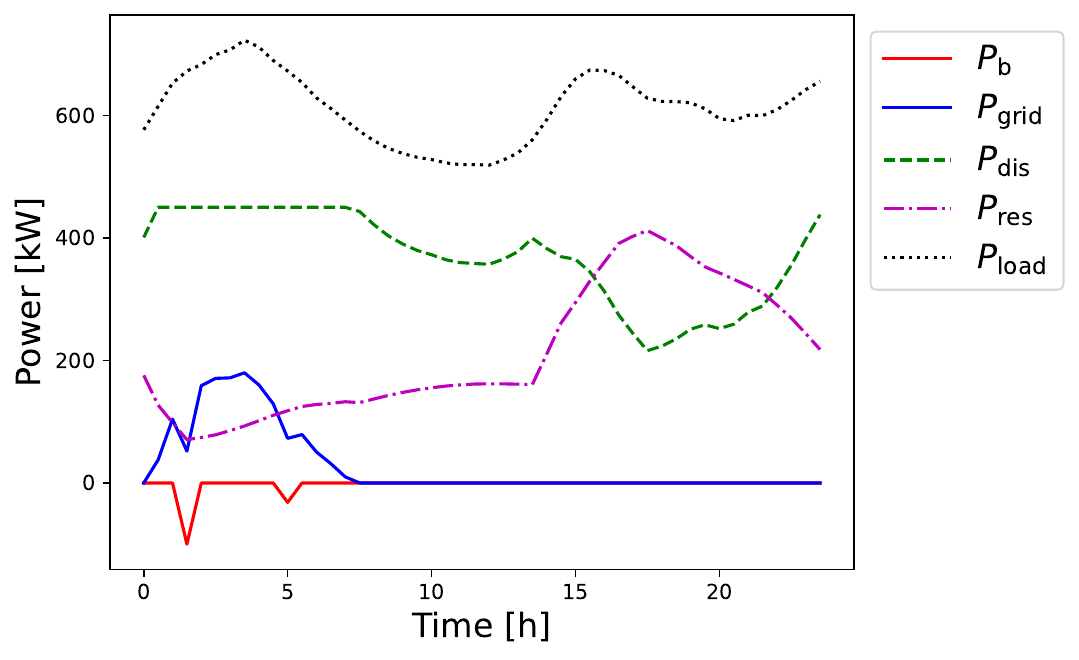}
    \caption{RL-based approach solution.}
    \label{subfig:RL}
    \end{subfigure}
    \caption{The figures above depict the solutions of the continuous variables for the optimal solution \ref{subfig:milp} and for the proposed approach \ref{subfig:RL} in a random time period of a $24$ hours selected from the testing dataset. The following continuous signals are shown: the power exchanged with the storage unit $P_\mathrm{b}$, the power exchanged between the microgrid and the main grid $P_\mathrm{grid}$, the total power generated by the microgrid $P_\mathrm{dis}$, the power generated by renewable energy sources $P_\mathrm{res}$, and the power demand of the microgrid $P_\mathrm{load}$.}
    \label{fig:power_solution}
\end{figure}

The reduction in computation time comes at the price of a small optimality gap -- under $1 \%$ -- and some instances of infeasibility, represented in Fig. \ref{fig:infeas_rate}.
The proposed approach consistently has significantly lower infeasibility rates than that of the supervised learning method across all the considered prediction horizons. The difference is more notable for $N_\mathrm{p}=48$, where the infeasibility rate for the proposed approach is $11/1000$ and that of the SL approach is $93.67/1000$. 
However, the supervised learning method has a slightly lower optimality gap than that of the proposed approach.
This reveals a trade-off between both learning approaches in our case study: while supervised learning favors optimality, reinforcement learning outperforms in terms of feasibility. Hence, the results suggest that the selection among these learning approaches ultimately depends on the relative importance of optimality and feasibility for a given application.

A simulation of the power exchanges in the online operation of the microgrid is shown in Fig. \ref{fig:power_solution}, where both the behavior of the optimal controller and of the RL-based controller are given. Regarding performance, the optimality gap between the optimal solution (MILP) and the RL-based approach is $0.35\%$.
The profiles of the continuous variables are almost identical for both the approaches, highlighting the RL-based method's ability to closely replicate optimal operation.
The associated costs for producing, selling, and buying energy are shown in Fig. \ref{fig:priceprofiles} in Appendix \ref{sec:sys_description}. 
For brevity, discrete variables are omitted, as they can be directly inferred from the continuous signals, e.g., $\delta_\mathrm{b}(k)=1 \iff P_\mathrm{b}(k) \geq 0$ and $\delta_\mathrm{grid}(k)=1 \iff P_\mathrm{grid}(k) \geq 0$.

Concerning the scalability for an increasing prediction horizon, as seen in Fig. \ref{fig:opt_gap}, the optimality of the proposed approach remains below $0.76\%$ and that of the SL-based approach remains under $0.18\%$, showing only small losses in optimality. The analysis of the infeasibility rate in Fig. \ref{fig:infeas_rate} reveals that the RL approach scales considerably better than the SL method regarding the infeasibility rate. The contrast is especially evident at $N_\mathrm{p}=48$, where the infeasibility rate of the proposed approach is $8.5$ smaller. As shown in Fig. \ref{fig:computation_time}, the reduction in computation time is consistent across the prediction horizons, achieving its peak at $N_\mathrm{p}=48$ where the maximum computation time is reduced by a factor of $16$. Finally, a box plot for the computation time for all approaches is shown in Fig. \ref{fig:boxplot}.

\subsection{Discussion}

The experiments presented in the case study in the previous section demonstrate the potential of the proposed framework for efficiently controlling MLD systems, achieving a favorable balance between computational time, optimality, and feasibility. This section brings to discussion several important aspects of the proposed approach.

Even though discrete actions $\epsd (k)$ that lead to the infeasibility of the LP \eqref{eq:lp} are discouraged, i.e., these actions have negative reward during training, the agent still may output an infeasible action. As feasibility is crucial in many applications, it can be restored either by an auxiliary control law, e.g., locally producing as much energy as possible and selling/buying the necessary power from the main grid, or by requesting a different action from the agent, e.g., selecting the second-highest Q-value instead of the highest. 
In the worst case, the mixed-integer linear program \eqref{eq:milp2} can be solved to optimality if there is sufficient time. Therefore, it is straightforward to incorporate the aforementioned fallback mechanisms to ensure feasibility in online operation.

The sampling time in the case study is relatively large compared to the average solution time of the MILP. However, the proposed approach offers advantages beyond the average reduction in computation time.
In control applications, it is crucial to ensure that the maximum solution time is predictable and remains within a reasonable bound, i.e.,
a solution must be computed before the next time step.
For a standard mixed-integer linear program solver, the worst-case complexity can be large compared to the average solution time, since, in the worst-case, all the possible combinations of discrete variables must be enumerated and solved for. 
In contrast, the proposed approach only requires evaluating a recurrent neural network and solving a linear program, both of which exhibit predictable computational behavior. 
Hence, it is relatively easier to estimate a reasonable worst-case bound on the computation time needed to determine the control input for the proposed approach. 
Due to the relatively smaller worst-case bound, 
the proposed approach requires the allocation of a shorter time window for computing the control signal.
As a consequence, the proposed approach can utilize more recent and accurate measurements of the system and exogenous signal, potentially improving the solution quality.
Finally, for larger systems with more decision variables and possibly nonlinear objectives or constraints, the disparity in the worst-case computation times is expected to increase, as the complexity of the optimization problem grows at each node of the branch-and-bound tree.

The proposed approach does not depend on a specific reinforcement learning (RL) algorithm. In this work, Deep Q-Learning was selected for its simplicity. Nonetheless, numerous algorithms, such as Rainbow DQN \cite{hessel2018rainbow} and PPO \cite{schulman2017proximal}, have shown better performance than that of Deep Q-Learning across several RL benchmarks. Therefore, it is likely that both optimality and feasibility rates could be further improved by employing more advanced RL algorithms.

A fundamental distinction between reinforcement learning and supervised learning lies in their objectives. Reinforcement learning seeks to maximize long-term rewards, which can be designed to minimize both operational costs and infeasibility rates. In contrast, supervised learning formulates a classification problem, aiming to establish a mapping from system states to optimal solutions. By its nature, reinforcement learning explores and evaluates infeasible actions, gaining insight into their consequences, whereas supervised learning focuses exclusively on predicting the optimal action. In the case study, the results suggest that reinforcement learning provides a richer output, offering more comprehensive information about actions and their feasibility. On the other hand, supervised learning is more accurate in predicting the optimal solution, but at the expense of a higher infeasibility rate. Further investigation is required to determine whether this observation extends to other problem domains.

In this work, modeling errors have not been considered. Nonetheless, the learning-based algorithms can be made more robust to model mismatch with techniques such as domain randomization \cite{tobin2017domain} and dataset augmentation via noise injection in the system state \cite{Goodfellow-et-al-2016}. 

The benefits of the learning-based methods are evident during online operation. However, training these methods requires accurate models and offline computational resources. 
For supervised learning, a dataset with the optimal solution of the system must first be obtained, after which the classifier weights are trained.
In contrast, for reinforcement learning, the agent repeatedly interacts with a simulated system to obtain data to simultaneously adjust the decoupled Q-function approximator -- possibly also a policy approximator -- via temporal-difference learning.
Fundamentally, with regard to training, the supervised learning approach requires the solution of mixed-integer programs, while its reinforcement learning counterpart requires the optimization of continuous decision variables. 
As a result, the computation cost in the reinforcement learning approach is smaller per iteration; however, it needs to perform more iterations due to exploration. 
Moreover, the determination of the best approach regarding offline computation load depends heavily on the type of available computational resources (e.g., CPU, GPU, and RAM)
as well as the specific implementation of the data acquisition and training algorithms, particularly the extent to which the algorithm is designed to exploit existing hardware. Given these factors, the comparison of the computation load required by the different methods is complex and beyond the scope of this work.

\section{CONCLUSIONS}\label{sec:conclusions}

This paper presented a novel approach that combines reinforcement learning and model predictive control for mixed-logical dynamical systems.
The online computation time of the MPC controller can  be significantly reduced by fixing the discrete decision variables with a policy trained by RL and then by optimizing over the continuous decision variables. This procedure effectively simplifies a mixed-integer linear program into a linear program.

For the training of the RL policy, we conceived the definition of the decoupled Q-function to decouple the discrete decision variables over the prediction horizon. Moreover, we showed how LSTM networks can be trained to approximate such a function.

Simulation experiments in a microgrid system revealed that our RL-based approach achieves a favorable trade-off between optimality, feasibility, and computation time. This suggests that the decoupled Q-function effectively captured the return of the sub-actions. Moreover, the computation time of the proposed approach was up to $16$ times faster than the standard MPC, whose solver was based on branch-and-bound. 

Several potential research directions for enhancing the performance of the proposed approach were discussed in the previous section. Additionally, the framework can be readily extended to control problems involving mixed-integer nonlinear programs, enabling its application to even more complex systems.

\section*{Data availability}

The code and data used in this paper are publicly available at \url{https://github.com/fabcaio/microgrid_RL}.



\section*{Appendix}

\begin{appendices}

\section{System description}\label{sec:sys_description}

In the case study, we address the economic dispatch problem in a microgrid system. This appendix describes the constituent elements of the microgrid, shown in Fig. \ref{fig:microgrid}, and their modeling as a mixed-logical dynamical (MLD) system. This modeling framework is used for its capacity to capture the behavior of continuous and discrete dynamics and decision variables. Besides, MLD modeling paves the way for the formulation of the economic dispatch problem as a mixed-integer program using an MPC approach. The use of MLD modeling and MPC for the microgrid operation optimization was explored in \cite{parisio2014model}, where the problem is cast as a mixed-integer linear program. The modeling of the microgrid was further simplified in \cite{pippia2019RuleBased} and \cite{masti19_warmStart} for the design of a ruled-based control policy and a learning-based control rule, respectively. Herein, we use the same simplified microgrid modeling framework to assess the performance of the proposed approach.

\subsection{Storage unit}

The energy storage unit is described by the following equations:
\begin{equation}\label{eq:ESS}
    x_\mathrm{b}(k+1) =
    \begin{dcases}
        x_\mathrm{b}(k) + \frac{T_\mathrm{s}}{\eta_\mathrm{d}} P_\mathrm{b}(k) & \mathrm{if} \ P_\mathrm{b}(k) < 0 \\
        x_\mathrm{b}(k) + T_\mathrm{s}\eta_\mathrm{c}P_\mathrm{b}(k) & \mathrm{if} \ P_\mathrm{b}(k) \geq 0
    \end{dcases}
\end{equation}%
where $x_\mathrm{b}(k)$ is the energy level in the storage unit at time step $k$, $\eta_\mathrm{c}$ and $\eta_\mathrm{d}$ are the charging and discharging efficiencies, $P_\mathrm{b}(k)$ is the power exchanged with the storage unit at time step $k$ and $T_\mathrm{s}$ is the sampling time of the discrete-time system. At a given time step, the storage unit can be either discharging or charging, depending on the sign of $P_\mathrm{b}(k)$.
In order to capture this hybrid behavior, we model the storage unit as a mixed-logical dynamical (MLD) system \cite{99bemporad}. A binary variable $\delta_\mathrm{b}(k)$ is introduced to signal whether the storage unit is charging or discharging, $\delta_b(k) = 1 \iff P_\mathrm{b}(k) \geq 0$ and $\delta_b(k) = 0 \iff P_\mathrm{b}(k) < 0$, respectively. By defining a continuous auxiliary variable $z_\mathrm{b}(k) = \delta_\mathrm{b}(k)P_\mathrm{b}(k)$, \eqref{eq:ESS} can be simplified and conveniently written as a single linear equation:
\begin{equation*}
    x_\mathrm{b}(k+1) = x_\mathrm{b}(k) + T_\mathrm{s}(\eta_\mathrm{c}-\frac{1}{\eta_\mathrm{d}})z_\mathrm{b}(k) + \frac{T_\mathrm{s}}{\eta_\mathrm{d}}P_\mathrm{b}(k)
\end{equation*}

\subsection{Generation units}

In the microgrid, energy can be locally produced either by dispatchable units or by renewable energy sources. The dispatchable units can be turned on/off, and their power output levels can be arbitrarily chosen by the microgrid operator within operating constraints. Concerning the renewable energy sources, solar and wind energy sources are considered.

The cost for locally producing energy at time step $k$ is
\begin{equation}\label{eq:cost_prod}
    C_\mathrm{prod}(k) = c_\mathrm{prod}(k)\sum_{i=1}^{N_\mathrm{gen}} P_i^{\mathrm{dis}}(k)
\end{equation}
where $c_\mathrm{prod}(k)$ is the cost of producing energy at time step $k$, $P_i^{\mathrm{dis}}(k)$ represents the power generated by dispatchable unit $i$ at time step $k$, and $N_\mathrm{gen}$ is the total number of dispatchable units. Besides, a binary variable $\delta_i^\mathrm{dis}(k)$ is introduced to represent whether the generator $i$ is turned on (1) or off (0). This will be useful in the next section when the operating constraints are introduced and the optimization problem is formulated.

Renewable energy sources are excluded from the expression \eqref{eq:cost_prod} because we assume zero cost for utilizing them when available. Moreover, due to their nature, the power generated by renewable sources is not controllable. The total power generated by renewable sources at time step $k$ is represented by $P_\mathrm{res}(k)$.

\subsection{Main grid}

At any given time step, the microgrid can buy or sell energy from the main grid. The power exchange at time step $k$ is represented by $P_\mathrm{grid}$(k). If this variable is nonnegative, the microgrid is set to import energy from the main grid. If it is negative, the microgrid is in export mode. Then, from the microgrid operator's perspective, at time step $k$, the operation cost is represented as
\begin{equation}\label{eq:cost_grid}
    C_\mathrm{grid}(k) =
    \begin{dcases}
        c_\mathrm{sell}(k) P_\mathrm{grid}(k) & \iff P_\mathrm{grid}(k) < 0 \\ c_\mathrm{buy}(k) P_\mathrm{grid}(k) & \iff P_\mathrm{grid}(k) \geq 0
    \end{dcases}
\end{equation}
where $c_\mathrm{sell}(k)$ and $c_\mathrm{buy}(k)$ are the prices for selling and buying energy to/from the main grid, respectively, at time step $k$.

Consider the discrete auxiliary variable $[\delta_\mathrm{grid}(k)=1] \iff [P_\mathrm{grid}(k) \geq 0]$ and the continuous auxiliary variable $z_\mathrm{grid}(k) = \delta_\mathrm{grid}(k)P_\mathrm{grid}(k)$. Now, with the use of MLD modeling, the operation cost between the microgrid and the main grid can be expressed in a single linear equation
\begin{equation}
    C_\mathrm{grid}(k) = c_\mathrm{buy}(k) z_\mathrm{grid}(k) - c_\mathrm{sell}(k) z_\mathrm{grid}(k) + c_\mathrm{sell}(k) P_\mathrm{grid}(k)
\end{equation}
Note that the operation cost is negative in export mode, i.e., the microgrid operator profits by selling energy to the grid.

\subsection{Assumptions}

We only consider uncontrollable loads, i.e., the microgrid operator does not affect the power demanded by them. Furthermore, at the current operation time, the actual load and its forecast over the future are assumed to be known. Similarly, the current power generated by renewable energy sources and its forecast are assumed to be known. These are not strong assumptions, since all of these values can be estimated with the use of historical data. We also assume knowledge of the market energy prices for buying and selling energy from the main grid and for locally producing energy with the dispatchable generators.

\subsection{Control scheme}

We consider a hierarchical control structure. A high-level controller is concerned with planning the operation schedule: i) which generation units are turned on and their power outputs; ii) whether the storage unit is charging or discharging and the corresponding power exchange; and iii) the mode of operation of the microgrid with respect to the main grid -- importing or exporting -- and the amount of power flowing between them. The variables determined by the high-level controller are reported in Table \ref{tab:decision_variables}. A low-level controller is responsible for keeping voltage, frequency, and phase within the operation range, and it operates in a faster timescale. In this scheme, the high-level controller is responsible for generating the set points for the microgrid elements so that the load demand is satisfied, and the operation cost is minimized. On the other hand, the low-level controller has the objective of tracking these set points computed by the high-level controller. In this work, we assume that the low-level controller is already in place, and we focus exclusively on the design of the high-level controller. 

\subsection{Control problem}

The operation cost is defined as the sum of the costs for locally producing energy and the costs for exchanging power with the main grid over a prediction horizon $N_\mathrm{p}$. Accordingly, the cost is defined as follows:
\begin{equation}\label{eq:cost}
    \begin{split}
        &J(\textbf{x}_\mathrm{b}(k), \gamma(k), \epsc (k), \epsd (k)) = \\
        \sum_{l=0}^{N_\mathrm{p}-1} &\left( C_\mathrm{prod}(k+l) + C_\mathrm{grid}(k+l) \right) = c^\T\epsilon_c(k)
    \end{split}
\end{equation}
where $\textbf{x}_\mathrm{b} (k) = [ x_\mathrm{b}^\T (0), \ldots, x_\mathrm{b}^\T (N) ]^\T$, $C_\mathrm{prod}(k)$ and $C_\mathrm{grid}(k)$ are defined in \eqref{eq:cost_grid} and \eqref{eq:cost_prod}, and $\epsc(k)$ is a stacked vector with the continuous decision variables over the prediction horizon. Furthermore, $c$ is a weighing vector that contains the market energy prices over the prediction horizon, and it is represented by
\begin{equation*}
    c = [c_0^\T,\ \ldots,\ c_{\mathrm{N_\mathrm{p}-1}}^\T]^\T
\end{equation*}%
where
\begin{equation*}
    c_k = [0,\ c_\mathrm{sell}(k),\ c_\mathrm{prod}(k)\cdot 1_{1\times N_\mathrm{gen}},\ 0,\ c_\mathrm{buy}(0) -c_\mathrm{sell}(k)]^\T
\end{equation*}%
\begin{figure}
    \centering
    \includegraphics[width=0.8\linewidth]{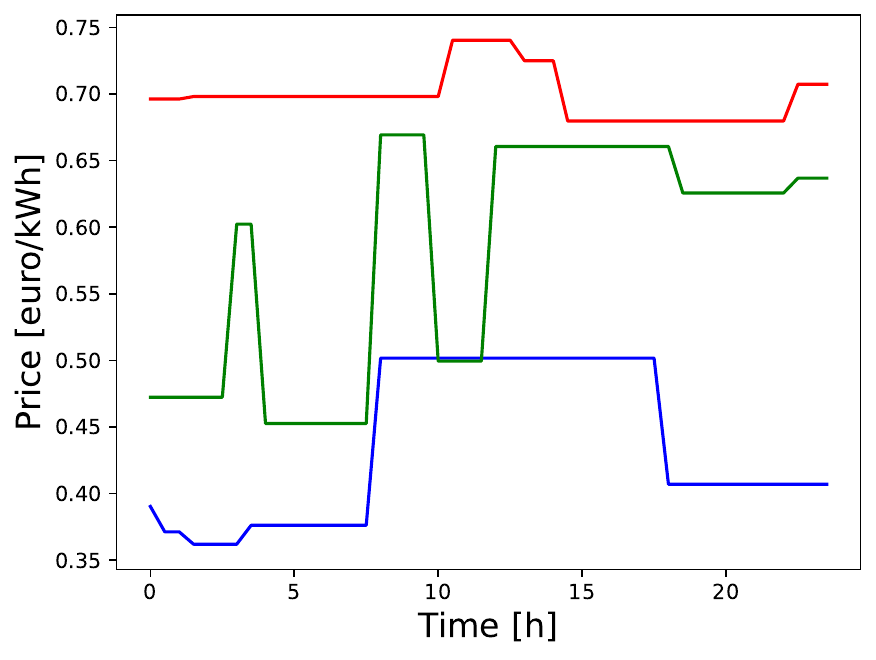}
    \caption{Price profiles for buying ($c_\mathrm{buy}$), selling ($c_\mathrm{sell}$), and producing ($c_\mathrm{prod}$) energy over a period of 24 hours.}
    \label{fig:priceprofiles}
\end{figure}

Note that the cost function $J(\cdot)$ is linear with respect to $\epsilon_\mathrm{c}(k)$ and that it can be negative in the scenario where the microgrid operator profits from selling energy to the main grid.
With the objective function defined, the optimization problem for the MPC controller can now be addressed. Along with minimizing the operation cost, the microgrid operator also has to satisfy some operating constraints, which are discussed next. Consider the following finite-horizon optimal control problem relative to the MPC controller:

\begin{subequations}\label{eq:mpc_migrogrid_full}
\begin{align}
    &\min_{\textbf{x}_\mathrm{b} (k), \epsc (k), \epsd (k)} \ J(\textbf{x}_\mathrm{b} (k), \gamma (k), \epsc (k), \epsd (k)) \nonumber \\
    \text{s.t. }
    &x_\mathrm{b}(k+l+1) = x_\mathrm{b}(k+l) + \label{eq:storage_unit} \\
    & \hspace{0.5cm}+ T_\mathrm{s}(\eta_\mathrm{c}-\frac{1}{\eta_\mathrm{d}})z_\mathrm{b}(k+l)  + \frac{T_\mathrm{s}}{\eta_\mathrm{d}}P_\mathrm{b}(k+l), \nonumber \\
    &E_2 \delta(k+l) + E_3 z(k+l) \leq E_1 u(k+l) + \label{eq:MLD_constraints} \\ 
    & \hspace{1cm} + E_4 x_\mathrm{b}(k) + E_5, \nonumber \\
    &P_\mathrm{b}(k+l) = \sum_{j=1}^{N_\mathrm{gen}} ( P_j^{\mathrm{dis}}(k+l) ) + \label{eq:cons_power_balance}  \\
    &+ P_\mathrm{res}(k+l) + P_\mathrm{grid}(k+l) - P_\mathrm{load}(k+l), \nonumber \\
    &\underline{P}_\mathrm{b} \leq P_\mathrm{b}(k+l) \leq \Bar{P}_\mathrm{b}, \label{eq:cons_power_battery} \\
    &\underline{P}_\mathrm{grid}(k+l) \leq P_\mathrm{grid}(k+l) \leq \Bar{P}_\mathrm{grid}, \label{eq:cons_power_grid} \\
    &\delta_j^\mathrm{dis}(k+l) \underline{P}_j^\mathrm{dis} \leq P_j^\mathrm{dis}(k+l) \leq \delta_j^\mathrm{dis}(k+l) \Bar{P}_j^\mathrm{dis}, \label{eq:cons_power_generated} \\
    & \text{for } j = 1,\ldots,N_\mathrm{gen},\ \text{and} \nonumber \\
    & \text{for } l = 0,\dots,N_\mathrm{p} - 1 \nonumber
\end{align}
\end{subequations}
where the cost $J(\cdot)$ is defined by \eqref{eq:cost} and the decision variables and parameters are described in Tables \ref{tab:decision_variables} and \ref{tab:parameters}. The first constraint \eqref{eq:storage_unit} concerns the dynamic of the storage unit. The equations \eqref{eq:MLD_constraints} arise from the introduction of the discrete and auxiliary variables introduced for the MLD modeling. The power balance of the microgrid and its connection with the main grid is enforced by constraint \eqref{eq:cons_power_balance} -- the generation must correspond to the load at all time steps. The lower and upper limits for the power exchange with the storage unit and with the main grid are represented in \eqref{eq:cons_power_battery} and \eqref{eq:cons_power_grid}, respectively. At last, the dispatchable generation units have minimum and maximum operation levels, which are represented by \eqref{eq:cons_power_generated}.

The objective function and the constraints of \eqref{eq:mpc_migrogrid_full} are linear with respect to the optimization variables. Since the model and the constraints are linear, they can be conveniently represented in MLD form, as in \eqref{eq:MLDsystem}. As a result, the optimization problem \eqref{eq:mpc_migrogrid_full} can be recast as the mixed-integer linear program \eqref{eq:milp2}.

\begin{table}[htb]
    \centering
    \fontsize{8pt}{8pt}\selectfont
    \resizebox{0.48\textwidth}{!}{%
    \begin{tabular}{|c|c|c|}
        \hline
         Parameter & Description & Value \\
         \hline
         $\Bar{x}_\mathrm{b}$ & maximum storage unit level & 250 [kWh] \\     $\underline{x}_\mathrm{b}$ & minimum storage unit level & 25 [kWh] \\
         $\Bar{P}_\mathrm{grid}$ & maximum power exchange with main grid & 1000 [kW] \\         $\underline{P}_\mathrm{grid}$ & minimum power exchange with main grid & -1000 [kW] \\
         $\Bar{P}_\mathrm{b}$ & maximum power exchange with storage unit & 100 [kW] \\         $\Bar{P}_\mathrm{b}$ & minimum power exchange with storage unit & -100 [kW] \\
         $\Bar{P}_i^\mathrm{dis}$ & maximum power output of dispatchable unit $i$ & 100 [kW] \\
         $\underline{P}_i^\mathrm{dis}$
         & minimum power output of dispatchable unit $i$ & 100 [kW] \\
         $\eta_c$ & charging efficiency of storage unit & 0.9 \\
         $\eta_d$ & discharging efficiency of storage unit & 0.9\\
         $N_{gen}$ & number of generators & 3 \\
         \hline
    \end{tabular}
    }
    \caption{Microgrid parameters, their descriptions, and their values.}
    \label{tab:parameters}
\end{table}

\section{Supervised learning approach}\label{sec:supervised_learning}

The training procedure of the SL approach presented in \cite{cauligi22_PRISM} is briefly summarized to highlight fundamental differences between both RL and SL approaches. The approach presented in \cite{cauligi22_PRISM} employs an LSTM network for classification, which is the same neural network architecture used in our RL approach for the policy approximation, facilitating the comparison between the performances of both methods. Let the optimal solution of the MPC problem in \eqref{eq:milp2} for a given state $\chi(k)$ be defined by the tuple $(\epsilon_{d}^*(k),\ \epsilon_{c}^*(k))$ containing the discrete and continuous optimization variables. The dataset for training is created by solving the MPC problem to optimality with branch-and-bound a number $N_\mathrm{data}$ of times with different initial augmented states, which are randomly sampled across the system's operating region of the state-space. The resulting pairs $\{(\chi(k),\ \epsilon_{d}^*(k))\}_{k=1}^{N_\mathrm{data}}$ are stored in a data buffer $\mathcal{D}$. Note that the continuous optimal solutions are not stored in the data buffer because the learning objective only concerns the prediction of the discrete optimal solutions. Having the dataset, it is straightforward to tune the parameters of the LSTM network by gradient descent on the following expression:
\begin{equation*}
	L = \frac{1}{T} \sum_{k=1}^{T}  \sum_{l=0}^{N_\mathrm{p}-1} g\left((y_{l}(k) - [\phi_\theta^{\mathrm{LSTM}}(\chi(k))]_l )\right)
\end{equation*}
where $g(\cdot)$ is a loss function, e.g., cross entropy function, $T$ is the size of the mini-batch, the target $y_{l}(k)$ is defined as the one-hot encoding of the optimal sub-action $\varepsilon^*_{\mathrm{d},l}(k)$ and the $l$-th output of the unrolled LSTM network with weights $\theta$ is denoted by $[\phi_\theta^{\mathrm{LSTM}}(\cdot)]_k$. For more details, the reader is referred to \cite{cauligi22_PRISM}.

\end{appendices}

\bibliographystyle{IEEEtran}
\bibliography{refs}

\begin{IEEEbiography}{Caio Fabio Oliveira da Silva}{\space} received the
B.Sc. degree from the University of Brasilia, Brazil, in 2017 and the M.Sc. degree from the Polytechnic University of Milan, Italy, in 2021. He is currently a Ph.D. candidate at the Delft Center for Systems and Control, Delft University of Technology, The Netherlands. His research interests include model predictive control, machine learning, and hybrid systems.

\end{IEEEbiography}

\begin{IEEEbiography}{Azita Dabiri}{\space}received the Ph.D. degree from the Automatic Control Group, Chalmers University of Technology, in 2016. She was a Post-Doctoral Researcher with the Department of Transport and Planning, Delft University of Technology, from 2017 to 2019. In 2019,
she received an ERCIM Fellowship and also a Marie Curie Individual Fellowship, which allowed her to perform research at the Norwegian University of Technology (NTNU), as a Post-Doctoral Researcher, from 2019 to 2020, before joining the Delft Center for Systems and Control, Delft University of Technology, as an Assistant Professor. Her research interests are in the areas of integration of model-based and learning-based control and its applications in transportation networks.
\end{IEEEbiography}

\begin{IEEEbiography}{Bart De Schutter}{\space} received the Ph.D. degree (summa cum laude) in applied sciences from KU Leuven, Belgium, in 1996. He is currently a Full Professor and Head of Department at the Delft Center for Systems and Control, Delft University of Technology, The Netherlands. His research interests include multi-level and multiagent control, model predictive control, learning-based control, and control of hybrid systems, with applications in intelligent transportation systems and smart energy systems. Prof. De Schutter is a Senior Editor of the IEEE Transactions on Intelligent Transportation Systems.
\end{IEEEbiography}

\end{document}